\documentclass[conference]{IEEEtran}
\usepackage{blindtext, graphicx}
\usepackage{parskip}

\let\sigproof\proof\let\proof\relax
\let\sigendproof\endproof\let\endproof\relax

\usepackage{amsthm}

\let\proof\sigproof
\let\endproof\sigendproof

\newtheoremstyle{sig}
  {}
  {}
  {\itshape}
  {}
  {\scshape}
  {.}
  {.5em}
  {#1 #2\thmnote{\quad(#3)}}

\theoremstyle{sig}

\newtheorem{dfn}{Heuristic}

\ifCLASSINFOpdf

\else

\fi

\begin{document}

\title{Conditions of Full Disclosure:\\The Blockchain Remuneration Model}

\author{\IEEEauthorblockN{S. Matthew English, Ehsan Nezhadian}
\IEEEauthorblockA{Fraunhofer-Institut f\"ur Intelligente Analyse und Informationssysteme IAIS\\
Rheinische Friedrich-Wilhelms-Universit{\"a}t Bonn\\
\texttt{\{english, nezhadian\}@uni-bonn.de}}}

\maketitle

\begin{abstract}
One of the fundamental applications for a practically useful system of money is remuneration. 
Information pertaining to the amount of compensation awarded to different individuals is often considered sensitive, commanding a certain degree of privacy. 
As Bitcoin and similarly designed cryptocurrencies evolve into a recognized medium of exchange for larger swaths of the world economy, an increasing number of people will earn income in the form of blockchain-based payments. 
The nature of these transactions is such that the minute details of an affected individuals compensation package and spending habits will be exposed to public scrutiny. 
In some cases this violates cultural norms which respect the confidentiality of salaries, yet in other cases it could be regarded as providing the benefits associated with greater transparency.
In this work we analyse the Bitcoin blockchain record of periodic payments accruing to an individual address in exchange for goods or services rendered.
For differing levels of available information we seek to determine the extent of insights that can be gleaned about the transacting counter-parties and the privacy implications this entails. 

\end{abstract}

\IEEEpeerreviewmaketitle

\section{Introduction}

All systems that preserve user privacy are alike; each system that violates user privacy does so in its own way. 
Those familiar with cryptocurrencies are aware that the mechanisms by which social beings mutually exchange value are constantly changing. 
Consider the modern word \textit{salary}, derived from the Latin \textit{salarium}, the root of which means \textit{salt}, an ancient medium of exchange.
To be ``\textit{worth one's salt}'' is an expression meaning that one contributes value in proportion to the amount one is paid. 
Under a regime of wages dispensed in salt there are doubtless a considerable set of challenges to overcome in the implementation and management of an efficient remuneration system. 
This work considers the problems with which we are confronted in the construction of systems build to compensate workers with a contemporary exchange medium, namely Bitcoin. 

In many cultures workers are uncomfortable disclosing their salaries publicly, yet for other professionals this information constitutes part of the public record.
Full disclosure of individual remuneration packages might violate personal privacy and erode the competitive advantage of an organization. 
However a policy of treating this data as openly available information is not without some potential benefit.
For instance the German public servant remuneration grade table (Tarifvertrag f{\"u}r den {\"O}ffentlichen Dienst der L{\"a}nder\footnote{\texttt{http://oeffentlicher-dienst.info/tv-l/}}) describes how research scientists, among others, are compensated for their labour.
Such open practices obviate the problems that arise from compensation differences between similarly skilled workers based on spurious criteria. 

In this work the authors were given access to an address on the Bitcoin blockchain, hereafter referred to as Alice\footnote{\texttt{1LWwLvKWbcpiZYqCcwfuQ3gjjNJkxftmEJ}}, which receives regular payments for goods and services rendered from an institution which we will denote Bob \& Company\footnote{\texttt{1EipJdYVJbqsTSQhj1icK424AkMbyjvgBm}}.
Through an analysis of the data presented in the sections to come, this work describes, as far as is known to the authors, the first empirical study of the privacy characteristics of live remuneration data extracted from the Bitcoin blockchain.
Insights discovered through the analysis of the data are presented.
Additionally we propose two heuristics for the identification of remuneration behaviours taking place on the Bitcoin blockchain.

\section{Obfuscation Techniques}

The poor privacy profile of blockchain-based currencies with design principles analogous to those of Bitcoin is well established \cite{miers2013zerocoin}. 
The most common mitigation to the risk posed by de-anonymization is to utilize a mix to shuffle bitcoins between different users. 
There are several of such services operating commercially and while specifics of the remedial measures vary slightly according to the provider there are some common drawbacks \cite{androulaki2013evaluating}.
These include the propensity of anonymity service providers to misappropriate funds, either explicitly or by going out of business.
As a result service providers typically offer short duration mixing which entails low transaction volumes and a restricted anonymity set. 
Additionally there is a further risk that the service provider will keep track of the coins they are ostensibly helping to anonymize and that eventually this information will be compromised.
Meiklejohn et al. \cite{meiklejohn2013fistful} have observed that among the mix services with whom they interacted, one simply stole their money while another twice returned the self-same coins, indicating that potentially they were, at the time, the only customers.
 
While mixing theoretically is a credible solution to the problem of how to preserve confidentiality of remuneration for employees compensated on the blockchain, in practice it poses a new set of difficulties and adds considerably to the complexity of the payroll process.
Consider that the business model of mix services necessitates users to pay a fee for each transaction involving coins anonymized.
Moreover the time and risk associated with such services constitute a cost over and above the explicit fee demanded for each mix. 
This serves as an impediment to organizations that seek to anonymize the transaction profile of their compensation activity.

As alluded to above, the public nature of the Bitcoin transaction log means that the use of pseudonyms is all that protects an individual organization or employee's privacy.
For those willing to consider cryptocurrencies other than Bitcoin two alternative approaches for the compensation of workers are Zerocoin and Zcash \cite{sasson2014zerocash}, both of which are blockchain-based cryptocurrencies that preserve the integrity of personal data in ways orthogonal to Bitcoin while posing their own unique set of risks. 

With issues of individual privacy there is often a trade-off between confidentiality and convenience, both of which can be evaluated on the basis of cost.
Organizations implementing business models that incorporate blockchain-based flows of valuable information need to determine the requisite level of privacy for their purposes and weigh this against the cost of taking the measures necessary to ensure that this level of privacy can be consistently achieved.

\begin{figure}[ht]
\includegraphics[width=0.5\textwidth]{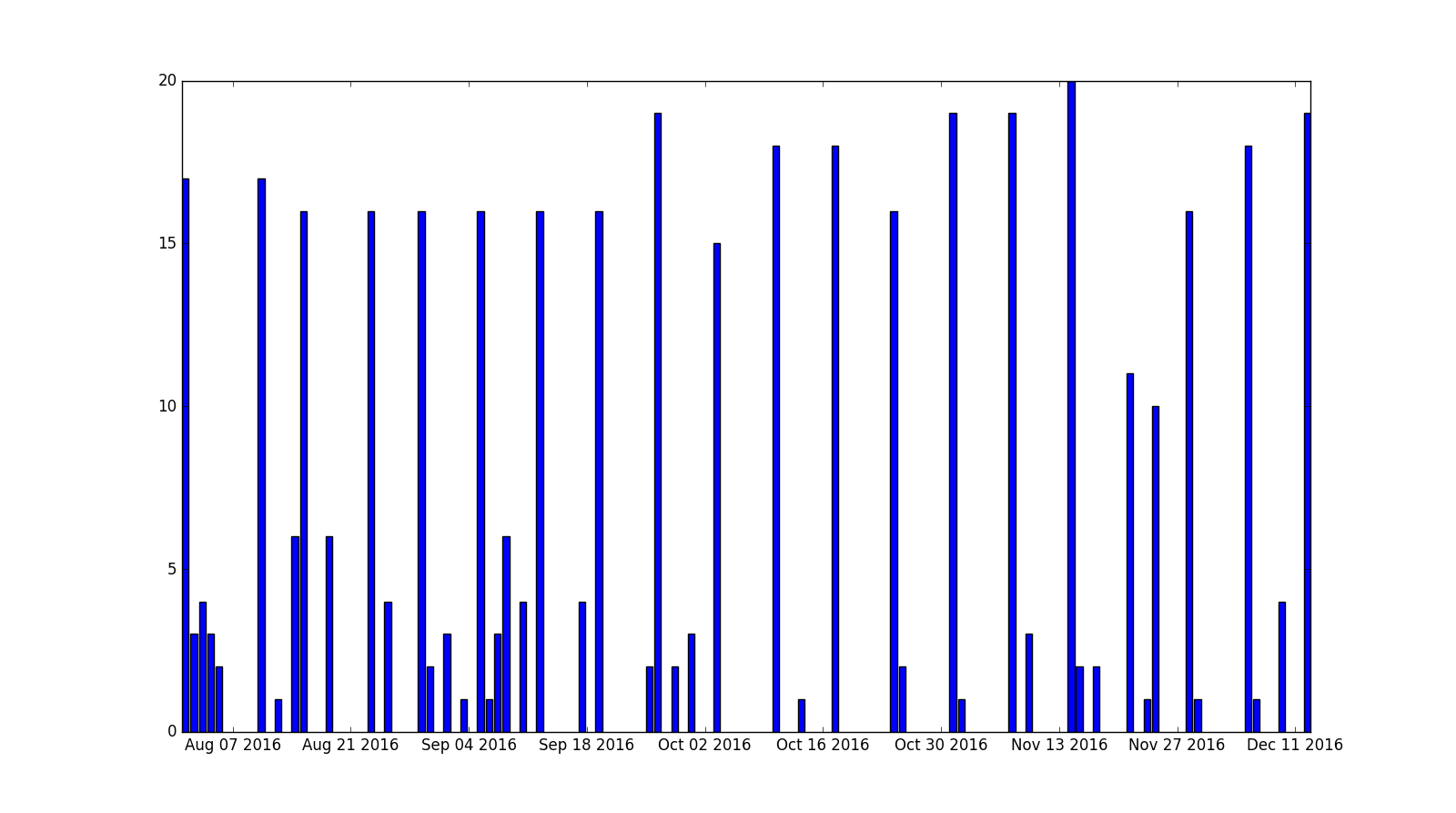}
\centering
\caption{Histogram of the 50 most recent transactions emanating from Bob \& Company. The y-axis represents the number of contributors, colleagues of Alice. The x-axis represents calendar days. The regular periodicity of payments to approximately the same number of addresses indicates behaviour characteristic of remuneration.}
\end{figure}

\section{Empirical Study}

The tendency for payments to accrue to Alice at regular intervals from the same address serves as a strong indicator that these transactions constitute remuneration.  

\begin{itemize}
    \item 2016-12-12 $\rightarrow$ 0.01413267 BTC 
    \item 2016-12-05 $\rightarrow$ 0.01731498 BTC 
    \item 2016-11-23 $\rightarrow$ 0.00272750 BTC
    \item 2016-11-14 $\rightarrow$ 0.03295339 BTC 
    \item 2016-11-07 $\rightarrow$ 0.00662214 BTC
    \item 2016-10-31 $\rightarrow$ 0.02340100 BTC
    \item 2016-10-17 $\rightarrow$ 0.00176991 BTC 
    \item 2016-10-10 $\rightarrow$ 0.00928812 BTC 
    \item 2016-10-03 $\rightarrow$ 0.01547708 BTC 
    \item 2016-09-26 $\rightarrow$ 0.01503248 BTC 
    \item 2016-09-19 $\rightarrow$ 0.02287458 BTC
    \item 2016-09-12 $\rightarrow$ 0.02176709 BTC 
    \item 2016-09-05 $\rightarrow$ 0.01249299 BTC 
    \item 2016-08-23 $\rightarrow$ 0.00240968 BTC 
    \item 2016-08-15 $\rightarrow$ 0.01973261 BTC
\end{itemize}  

The single address associated with Alice formed the initial basis of our analysis.
With this information alone it is trivial to observe that Bob \& Company is the sole source of all the transactions listed above. 
By exploring the transaction profile associated with Bob \& Company we generated the histogram in Figure 1, this information yields insight into (at least) a consistent subset of the number of accounts payable (presumably employees) with whom Bob \& Company has regularly interacted over the course of the time period depicted. 
Based on the behaviour exhibited by Bob \& Company as described by Figure 1 we catalogue a heuristic for the identification of organizations compensating employees over the Bitcoin blockchain.
$\newline$
\begin{dfn}[Remuneration Profile]
Whereas address $\alpha_o$ transmits payments to disparate addresses ($\alpha_1$, $\alpha_{\ldots}$, $\alpha_n$), for amounts consistently in the range $\rho$, at intervals $\iota$, with regularity $\iota_\tau$ $\Rightarrow$ $\alpha_o$ is affiliated with an organization of which $\alpha_1$, $\alpha_{\ldots}$, $\alpha_n$ are employees.
\end{dfn}

Assignment of Heuristic 1 is complicated by the fact that over the lifetime of the address affiliated with Bob \& Company we can observe three distinct patterns of behaviour, the first two of which were not readily identifiable as constituting a profile attributable to the remuneration of employees. 
This can be discerned through careful examination of Figure 2 wherein the first phase is characterized by largely random activity, followed by a period of quiescence, and finally the structured organizational behaviour described by Figure 1.
The period of activity described by Heuristic 1 is more recent than the anomalous periods, therefore we weight it as more significant to our analysis.

\begin{figure}[h]
\includegraphics[width=0.5\textwidth]{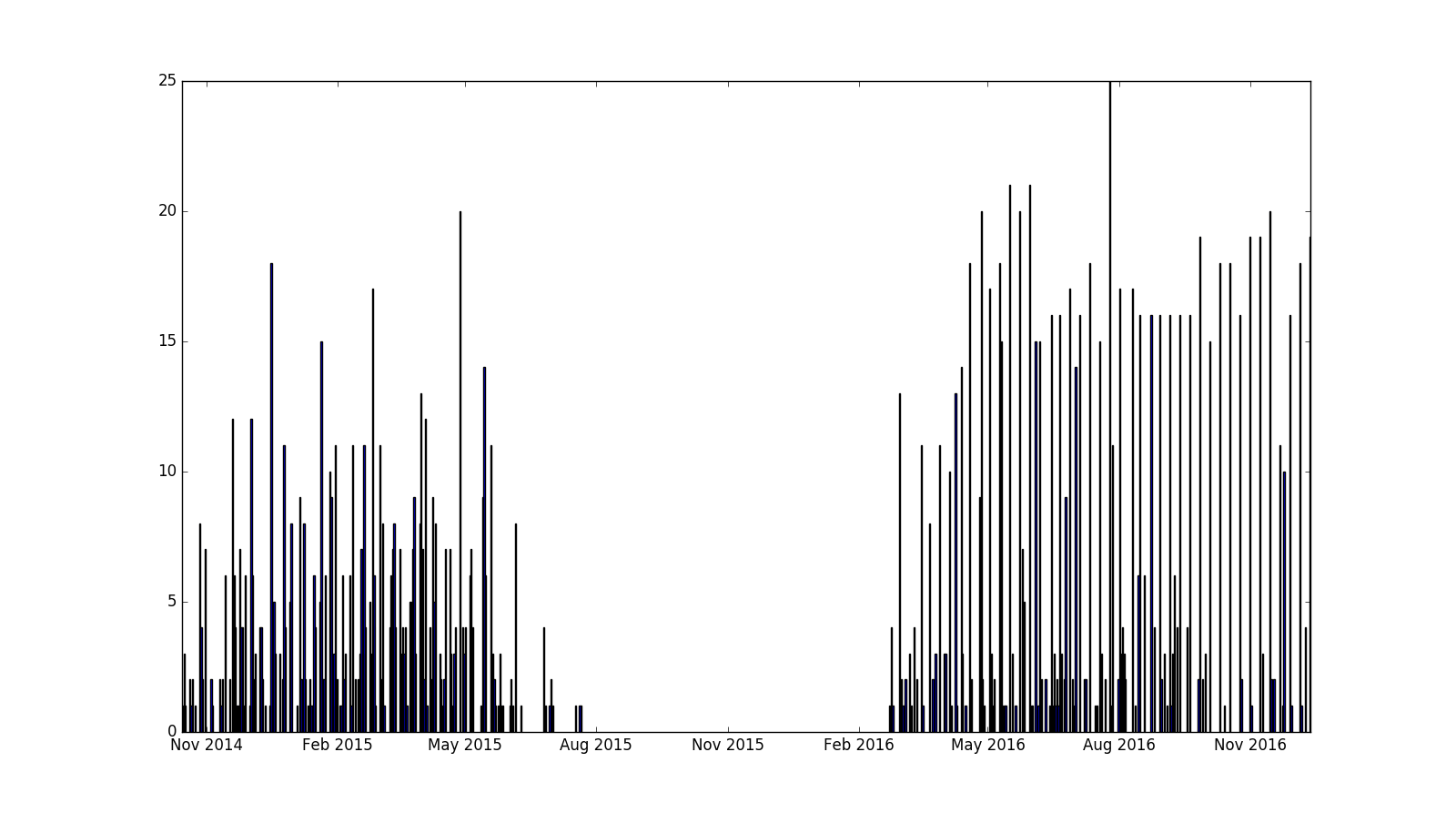}
\centering
\caption{The complete lifetime histogram of Bob \& Company. The y-axis represents number of addresses to whom transactions are broadcast, the x-axis represents time.}
\end{figure}

Best privacy practices of Bitcoin recommend that a ``fresh'', never before used, address be utilized for each new transaction. 
It is clear that neither Alice nor Bob \& Company follow this procedure.
This fact notwithstanding, the majority of transactions transmitted by Bob \& Company do in fact appear to follow the one address per transaction protocol. 
In Figure 3 we can see a comprehensive break-down of the transactions made by Bob \& Company to distinct addresses. 
The Green section (64.38\%), the majority of cases, represents transmission to an address that has never before been issued to by Bob \& Company and is never issued to again. 

\begin{figure}[ht]
\includegraphics[width=0.5\textwidth]{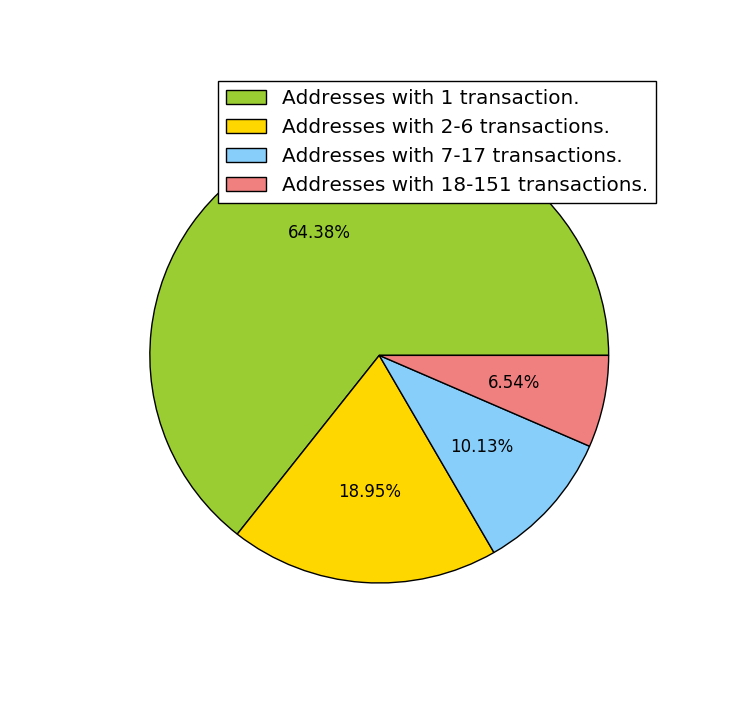}
\centering
\caption{The proportion of transactions partitioned according to the number of times Bob \& Company transmitted Bitcoin to the address in question.}
\end{figure}

Generally the Bitcoin transmissions from Bob \& Company occur at consistent intervals, categorized by a certain regularity in time. 
While this is true for the majority of addresses to whom Bob \& Company transmit it is not a universal rule, this is observable in the discrepancy between Figure 4, a representative example of (temporally) regular Bitcoin transmissions, and Figure 5, which depicts (atypically for Bob \& Company) a highly irregular pattern of behaviour. 

\begin{figure}[h]
\includegraphics[width=0.5\textwidth]{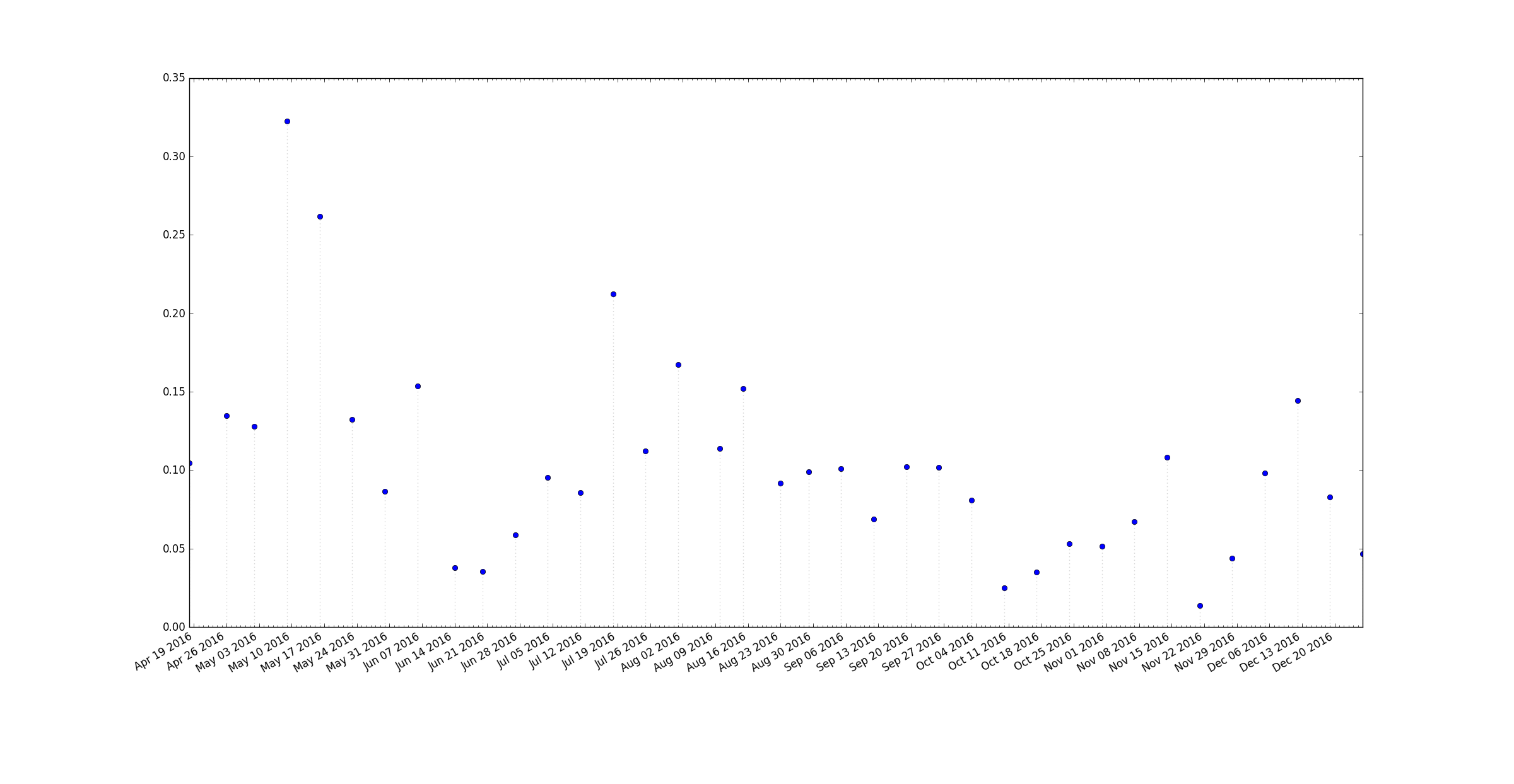}
\centering
\caption{Prototypical example of transactions from Bob \& Company to an address, note the characteristic regularity. The x-axis is the amount in Bitcoin, the y-axis represents time.}
\end{figure}

\begin{figure}[h]
\includegraphics[width=0.5\textwidth]{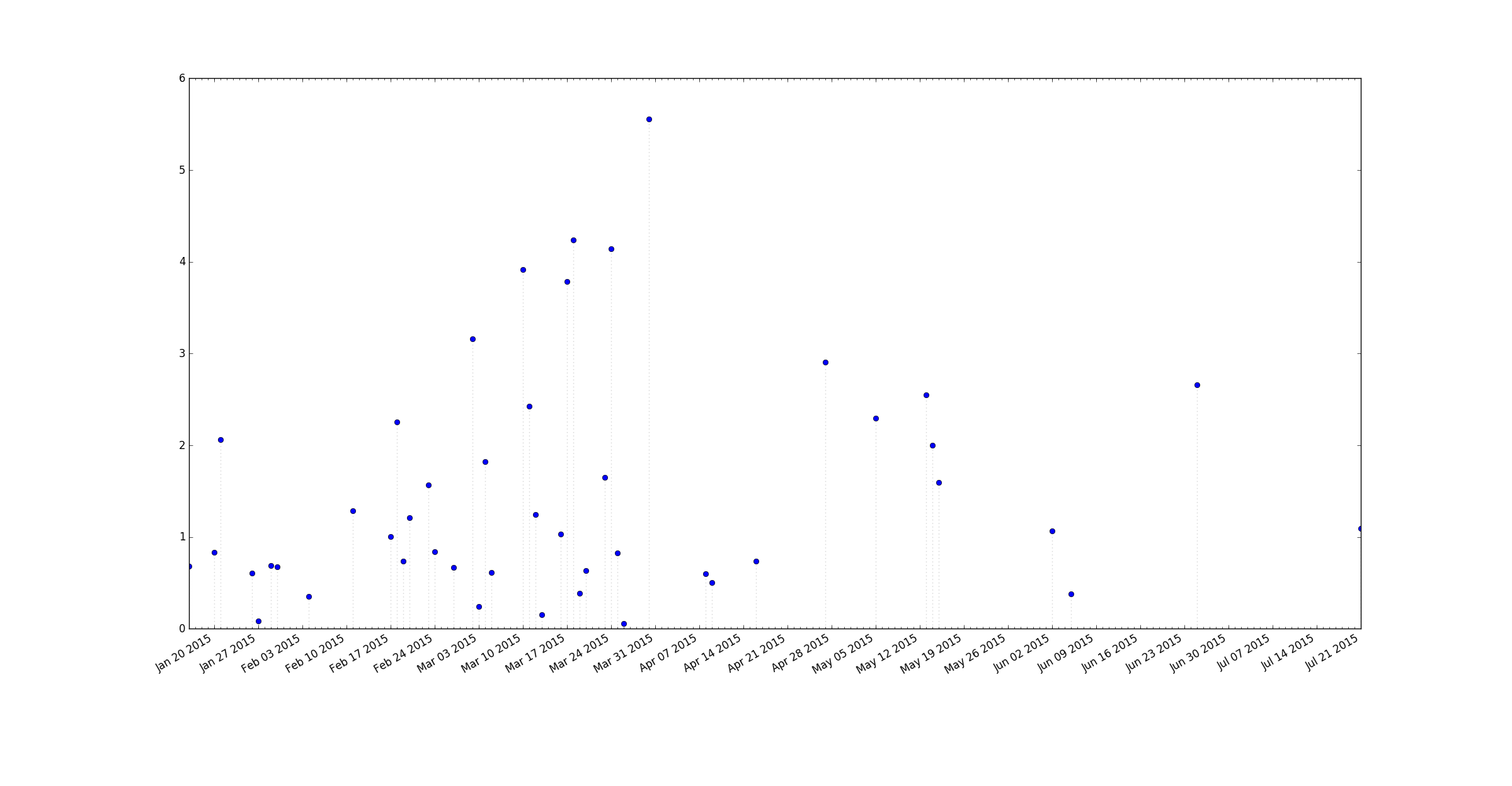}
\centering
\caption{Observe that although the majority of transactions emanating from Bob \& Company are quite regular there are still those that are described by highly erratic activity as we see here. The x-axis is the amount in Bitcoin, the y-axis represents time.}
\end{figure}

\subsection{Volatility in Amount of Bitcoin Transmitted}

In stark contrast to the regularity in terms of periodicity of transactions by Bob \& Company is the irregularity in the amount of Bitcoin involved in each transaction. 
One of the reasons for this is that compensation to employees will fluctuate with the value of Bitcoin. 
As such if transactions closely track a consistent benchmark or reference point in a national currency, such as the United States dollar, it is another strong indicator that they represent salary payments. 

\begin{figure}[h]
\includegraphics[width=0.5\textwidth]{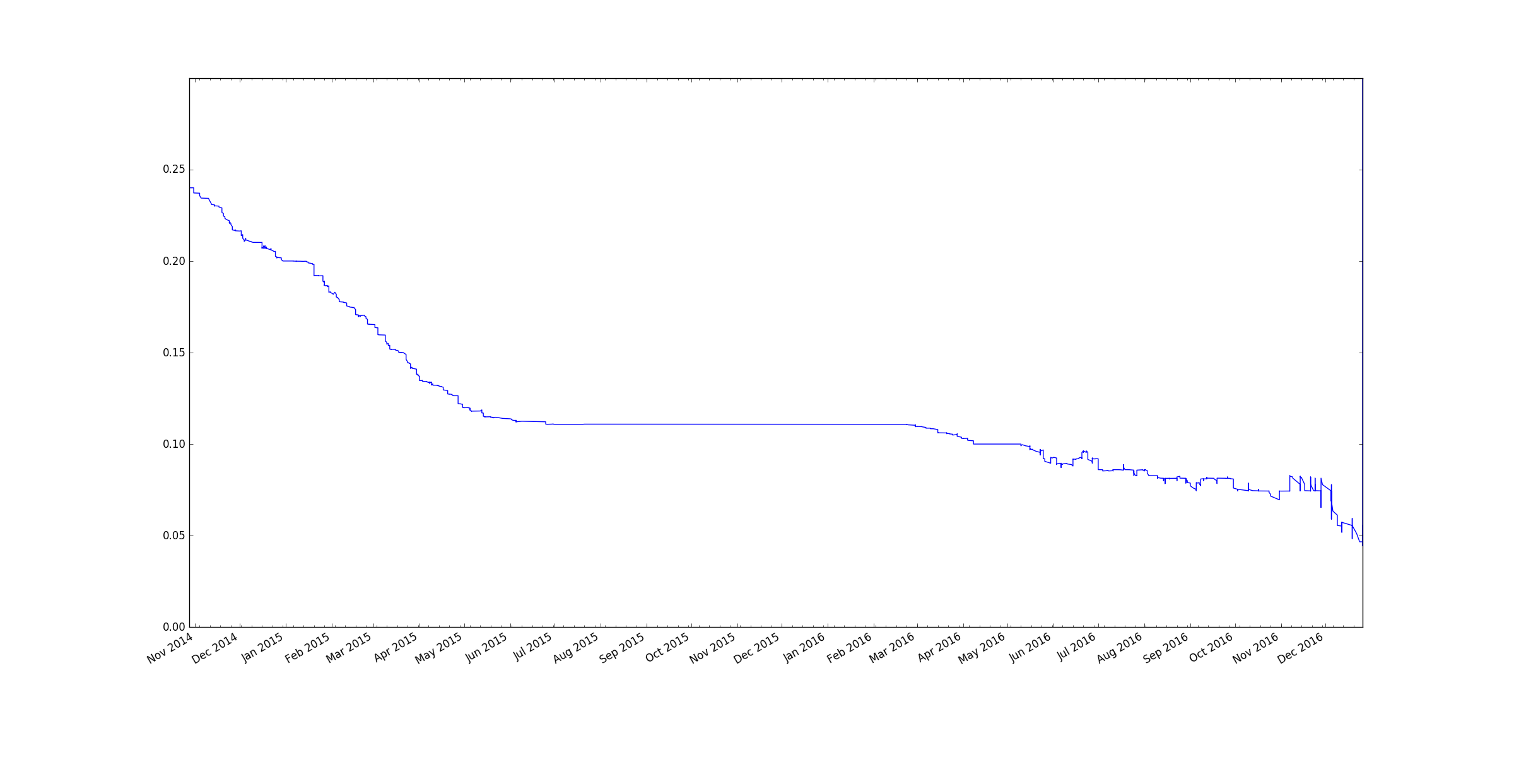}
\centering
\caption{The median transaction value in Bitcoin sent by Bob \& Company over the lifetime of the address. Note that as the price of Bitcoin increases relative to the United States dollar the value of transactions emanating from Bob \& Company is decreasing in rough proportion.}
\end{figure}

\begin{dfn}[Benchmark Target]
If the median $\mathcal{M}$ of payments transmitted $\mathcal{P_T}$ by address $\alpha_o$ to disparate addresses ($\alpha_1$, $\alpha_{\ldots}$, $\alpha_n$), consistently tracks a target value of national currency $\mathcal{N}$, across various exchange rate ($\mathcal{E}$) climates $\Rightarrow$ $\mathcal{P_T}$ are remuneration for goods or services rendered. 
\end{dfn}

In November of 2014 when the price of one Bitcoin was approximately USD \$340.00 the average compensation from Bob \& Company was 0.25 Bitcoin per transaction. 
In December of 2016 when one Bitcoin typically sold for a price of approximately USD \$930.00, the average compensation from Bob \& Company was 0.05 Bitcoin per transaction.
Therefore, from the period during which data is first available for Bob \& Company until the time of writing, the price of Bitcoin has increased and the amount of remuneration from Bob \& Company have decreased in proportion to each other as described by Figure 6.

\section{Cost Calculation}

The practical considerations involved in deploying mixing services and the degree of scale necessary to meet large payroll obligations are multitudinous. 
Organizations and institutions that need to regularly transmit large payments make themselves susceptible to attacks well known to the communication mixing community, e.g. \textit{packet counting}, or the \textit{intersection} attack. 
Even if one were to employ the services of a mix, with the risks that this entails, one is still left exposed to side channel attacks from an adversary that is sufficiently committed.

\section{Implications}

The practice of compensating employees via blockchain-based cryptocurrencies has made possible the realization of heretofore unimagined business models.
Bitcoin facilitates the compensation of contributing individuals across the world regardless of whether or not they have a consistent home address or a bank account. 
The practice of paying salaries with this contemporary medium of exchange is growing in, if not popularity, at least notoriety, as indicated by corporations ranging from machine learning focused hedge funds to small start-ups proclaiming their affinity for remunerating their workers using Bitcoin \cite{Faceless}. 
In this work we have demonstrated the substantive privacy concerns raised by the practice of awarding salaries using cryptocurrencies with design principles similar to those of Bitcoin by the meticulous analysis of live blockchain data.
In this section we explore some of the implications of the possibilities unleashed by this mechanism of disseminating personal salaries. 

\subsection{Industrial Espionage}

In this work we were able to track the growth of Bob \& Company. 
If this analysis were undertaken by a competitor of Bob \& Company it could erode their competitive advantage by divulging information relating directly to the economic viability, growth patterns and trajectory of Bob \& Company's business. 

\subsection{Endangering Employees}

The Women's Annex Foundation (WAF)\footnote{\texttt{http://digitalcitizenfund.org/}} encourages girls in Afghanistan to engage in blog writing, software development, video production and social media marketing, paying them for their efforts in Bitcoin.
The heuristics described in this work could be used to identify organizations on the blockchain, organization like the WAF. 
Business models with similar objectives do well to consider whether compensating workers via the blockchain is consistent with promoting the well being of their contributors and if decided in the affirmative, to take all necessary precautions to sufficiently anonymize their transaction profile. 

\subsection{Corporate Governance}

While doubtless there are ills associated with increased transparency there are also considerable benefits. 
Increased oversight, transparency, and participation on behalf of stake-holders is realizable as never before through the deployment of public ledger based value transmission systems. 
This could herald a new ethos in corporate governance. 

\subsection{Open Budget Initiative}
There are at the moment projects underway from various governments and civic institutions, e.g. the World Bank Institute, to promote the kind of budget transparency that can decrease corruption and improve living standards, the kind of transparency detailed in this work. 

%
%
%

\section{Conclusion}

In the early days of Bitcoin the perceived anonymity of this value transfer technology was one of it's most attractive features, helping to fuel it's adoption on marketplaces such as the Silk Road. 
Today it is clear that the anonymity guarantees of Bitcoin are tenuous. 
This work provides a foundation for the creation of mechanisms that might search the blockchain for evidence of remuneration behaviour taking place using cryptocurrencies, i.e. Bitcoin. 
In consideration of the ethics of anonymity we do well not to overlook the multitude of important reasons for anonymity that we might take for granted with traditional currencies.
It is still the case that many people are uncomfortable divulging the details of their salaries with friends or coworkers. 
The relative ease with which individual addresses in the Bitcoin blockchain can be associated with a salary through the heuristics herein presented demonstrates a host of new challenges and opportunities. 
This work represents a first step in the determination of what this paradigm shift will ultimately have in store for the way we relate and interact with one another through one of the oldest social technologies, money.

\section*{Disclaimer}

The authors of this work obtained explicit permission from the owners of the addresses under direct consideration to conduct the empirical study herein presented.

\ifCLASSOPTIONcaptionsoff
  \newpage
\fi

\begin{IEEEbiography}[{\includegraphics[width=1in,height=1.25in,clip,keepaspectratio]{picture}}]{John Doe}
\blindtext
\end{IEEEbiography}

\end{document}